\documentclass[prl,aps,floats,twocolumn,tightenlines,notitlepage,preprintnumbers,showpacs]{revtex4}
\usepackage{epsfig}
\usepackage{amsmath}
\usepackage{graphics}
\usepackage{epsfig}

\newcommand{\bea}{\begin{eqnarray}}
\newcommand{\eea}{\end{eqnarray}}
\newcommand{\beq}{\begin{equation}}
\newcommand{\eeq}{\end{equation}}

\begin{document}
%%%%%%%%%%%%%%%%%%%%%%%%%%%%%%%%%%%%%%%%%%%%%
\title{Escape of black holes from the brane}
\author{Antonino Flachi}
\email{flachi@yukawa.kyoto-u.ac.jp}
\affiliation{Yukawa Institute for Theoretical Physics, Kyoto University, Kyoto 606-8502, Japan}
\author{Takahiro Tanaka}
\email{tama@scphys.kyoto-u.ac.jp}
\affiliation{Department of Physics, Kyoto University, Kyoto 606-8502, Japan}
%\date{May 2005}
\preprint{YITP-05-28}
\preprint{KUNS-1974}
\begin{abstract}
TeV-scale gravity theories allow the possibility of producing small
 black holes at energies that soon will be explored at the LHC or at
 the Auger observatory. 
One of the expected signatures is the detection of Hawking radiation,
 that might eventually terminate if the black hole, once perturbed,
 leaves the brane. 
Here, we study how the `black hole plus brane' system evolves once the
 black hole is given an initial velocity, that mimics, for
 instance, the recoil due to the
emission of a graviton. The results of our dynamical analysis 
 show that the brane bends around the black hole,
 suggesting that the
black hole eventually escapes into the extra dimensions once two
 portions of the brane come in contact and reconnect. This gives a
 dynamical mechanism for the creation of baby 
branes.
\end{abstract}
\maketitle
%%%%%%%%%%%%%%%%%%%%%%%%%%%%%%%%%%%%%%%%%%%%%%
\noindent
{\it Introduction}:
The reason why gravity is so much weaker than all the other forces,
%$G_N/G_F \sim 10^{-33}$, 
aside from being still a mystery ({\it the
hierarchy problem}), constitutes 
the major obstacle in performing experiments in the realm of quantum
gravity, for the rather obvious reason that overcoming the fundamental
Planck scale requires center of 
mass energies greater than $10^{18}$ GeV. 

This common belief was radically questioned a few years ago. 
Although in four dimensions there is no hope to reach such an energy 
in a terrestrial experiment, the situation
drastically changes when one assumes the existence of extra dimensions. 
In the latter case it might even be possible to reach a transplanckian
regime. When geometry and scales of the extra (space-like) dimensions
are appropriately chosen, the higher dimensional Planck scale may turn
out to be much smaller than $10^{18}$ GeV, thus reformulating the 
hierarchy problem in geometrical terms.

The details depend upon the specific realisation of the model, two popular examples
being scenarios with large \cite{add} or warped extra dimensions \cite{rs}, but the
common features are a low fundamental Planck scale, $M\sim$TeV, the localisation of the
standard model on branes, representing our directly observable universe, and the
propagation of gravity throughout the higher dimensional space. This is, roughly, what
is known as the brane world.

In the past few years many people started to use these simple ideas to
build models of various type ranging from cosmology to particle physics
and, more speculatively,
investigate how the brane world can be used as a tool to solve long
standing problems in physics. To date no complete solution to any
problem has yet been found and
many of the predictions of the brane world, relying on tunings of various
sorts, do not provide definitive and clear-cut answers. However, it is
widely believed that, if the 
fundamental scale of gravity truly lies in the TeV range, a very
spectacular and relatively model independent prediction can be made,
that is the creation of small black 
holes in the high energy collision of two particles \cite{dl,gt}. Thus
the LHC or the Auger Observatory might be able to perform quantum
gravity experiments and initiate the study of 
black hole microphysics.

A generic assumption is to consider the production of 
black holes whose mass exceeds the fundamental Planck scale $M$. In this
regime a semiclassical treatment is possible and quantum
gravity corrections can be ignored. The size of the black holes produced
at colliders is typically assumed to  be much smaller than the
characteristic length of the extra 
dimensions. Then it seems reasonable to describe these objects by higher
dimensional asymptotically flat solutions \cite{tangherlini,mp}. In
this approximation, the Schwarzschild radius %of $R_s$ 
of a $d$-dimensional black hole of mass $m$ is given by 
\beq
R_s = {1\over \sqrt{\pi}}\left({8\Gamma((d-1)/2)\over d-2}\right)^{1\over(d-3)}
\left({m\over M}\right)^{1\over(d-3)} {1\over M}~.
\nonumber
\eeq
We use units in which $c=\hbar=1$. 
Given these assumptions, one considers two particles with center of mass energy
$\sqrt{s}$ moving in opposite directions at impact parameter $b$. When $b$ is smaller than
$R_s$ with $m$ replaced by $\sqrt{s}$, 
semiclassical arguments show that a marginally trapped surface forms at the overlap
between the two colliding shock waves describing the two scattering
particles. This implies the formation of a common horizon unless naked 
singularities are formed. 
The black disk approximation, then, suggests the following formula for the cross section:
\beq
\sigma_{BH} \sim \pi R_s^2~.
\nonumber
\eeq
Assuming a higher dimensional Planck scale $M \leq 3$TeV, 
the cross section will range between $10^{-2}$ to $10^2$ pb 
at the LHC energy, $\sqrt{s} \sim 14$ TeV. 
At the luminosity $L \sim 10^{34}$ cm$^{-2}$ s$^{-1}$, the LHC will be
able to produce about $10^{7}$ black holes per year.

After the black hole is formed, it will decay by Hawking radiation at a temperature
\beq
T_H = {d-3\over 4\pi R_s}~.
\nonumber
\eeq 
As it was argued in Ref.~\cite{ehm}, Hawking evaporation must emit comparable
amounts of energy into each low energy effective degree of freedom 
in the bulk and on the brane. Then, if we assume that only gravity propagates in
the bulk, radiation on the brane will be the dominant component of the Hawking radiation.

However, this is not the whole story. As noticed in Ref.~\cite{frolov}, the black hole
is expected to be rotating and thus exhibits superradiance; this would enhance the
emission of higher spin particles, possibly making the emission of gravitons a dominant
effect, thus strongly perturbing the system and eventually resulting in the black
hole leaving the brane with sudden termination of the Hawking
radiation. Ref.~\cite{frolov} discusses this on the basis of a field theory model,
where the black hole,
treated as a point radiator, is described as a massive scalar field with internal
degrees of freedom. Ref.~\cite{stojkovic} pushes this further and argues that it is
possible to distinguish between ${\mathcal Z}_2-$ and non ${\mathcal Z}_2-$symmetric
scenarios due to the fact that a black hole cannot recoil if the spacetime is
${\mathcal Z}_2-$symmetric. 

In this paper we would like to reconsider this problem. Although the interaction
between defects and black holes is a relatively well studied subject \cite{cfl}, most
of the investigations performed so far focus on the static case. 
Apart from this, it is assumed 
that a black hole may leave the brane somehow, but 
there is no direct example showing how such departure may occur. In the
present work we will look at the motion of a brane in the gravitational
field of a (small) black 
hole, and see explicitly how a
black hole can escape from the brane. 
% by solving the dynamical problem, and . 
%As a result of our investigation all the quantities of interest can
%be directly evaluated and the various regimes studied directly.

As we will see, due to the presence of the black hole, even in the most crude
approximation, any perturbation that will give the brane an initial velocity with
respect to the black hole, will cause deformations in the brane itself. This is
essential to understand whether or not a black hole can leave the brane. We will also
see that these deformations induced in the brane can be indeed simulated in a precise
way, allowing us to determine the time scale, $\tau_*$, at which this separation
eventually occurs and to compare it with the lifetime $\tau$ of the black hole. 
This will tell us whether or not the escape occurs before the 
black hole evaporation completes.

\noindent
{\it Membrane dynamics}:
We shall describe the set-up and the limitation of our approach.
As we have mentioned, the question we would like to answer is whether or not a black
hole on a brane leaves the brane once the system is
perturbed and, if this is the case, work out the time scale at which this process occurs. 

We will consider a general $d-$dimensional bulk spacetime and assume that our universe is 
a $(p+1)-$dimensional brane. We consider black holes whose gravitational
radius is much smaller than the size of the extra dimensions, and 
this allows us to assume that our spacetime is adequately described by the
asymptotically flat solution \cite{tangherlini, mp}. This is of course a
reasonable assumption only if the self-gravity of the 
brane is negligible, which is  the case when the size of black
holes is sufficiently small. 
%{\it i.e.} if the tension is very small and, 
As long as this approximation is reasonable, our
results will give a model-independent prediction. We assume that the
black hole is formed out of matter on the brane, and symmetries require
that the brane initially lies on the equatorial plane of the black hole. 
From the point of view of accelerator generated black holes, the interesting
possibility is to consider rotating black
holes, however here we start with the case of Schwarzschild black holes
as a first step, hoping to capture the essential features of the
problem, %and in order to compare
and deferring the rotating case to our forthcoming work
\cite{nino-takahiro-forth}. Thus the bulk space is
described by the following line element:
\beq
ds^2 = -f(r) dt^2 + f^{-1}(r) dr^2 + r^2 d \Omega^2_{d-2}~,
\nonumber
\eeq
where the function $f(r)=1-(1/r)^{d-3}$. We set the horizon 
radius to unity by adjusting the unit of the length. 
Then the area of the event horizon is equal to that of a unit $(d-2)-$sphere,
$\Omega_{d-2}$. 

The brane in the leading approximation is described by a Dirac-Nambu-Goto action:
\beq
S = - \sigma \int d^{p+1}\zeta \sqrt{\gamma}~,
\nonumber
\eeq
where $\sigma$ is the tension of the brane and $\gamma$ is 
the determinant
of the induced metric $\gamma_{ij}$ on the brane. 

\begin{figure}
\scalebox{0.5} {\includegraphics{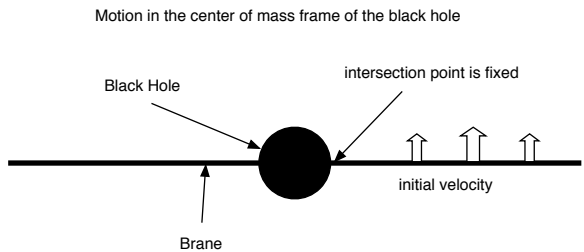}}
\caption{Schematic picture of the system in the center of mass frame of
 the black hole. The intersection point is held fixed and the brane is
 given an initial velocity.}
\label{motion}
\end{figure}

We use $\{\zeta^a\}\equiv
\{t,r,\mbox{\boldmath$\chi$}\}$ with $a= 0,1, ..., p$ 
as the coordinates on the brane, where 
{\boldmath$\chi$} represent 
coordinates of a $(p-1)$-dimensional sphere. 
Trajectory of a spherically symmetric brane 
is specified by 
the azimuthal inclination angle $\theta(t,r)$. 
%\begin{equation}
%x^A = x^A(\zeta^a)
%\end{equation}
%where $A= 0, 1, ..., d$ and 
%${\zeta^a}$
%In the vicinity of the membrane a convenient choice of coordinates is $(\zeta^a, z)$ with the $z-$direction being orthogonal to the wall. In these coordinates the invariant line element has the form
%\begin{equation}
%ds^2 = g_{AB}dx^Adx^B = g_{AB}x^A_{,a} x^B_{,b} d\zeta^a d\zeta^b -dz^2~,
%\end{equation}
%and 
The induced metric on the brane is then given by 
\begin{eqnarray}
\gamma_{ab} d\zeta^a d\zeta^b &= & 
  - f(r) dt^2 + f^{-1}(r)dr^2 \cr
   && +r^2 
  \left[(\theta_t dt +\theta_r dr)^2+\sin^2\theta d\Omega_{p-1}^2\right]~,  
\nonumber
\end{eqnarray}
where 
the notation $h_x \equiv \partial h /\partial x$ is used, 
and we have 
\beq
S_{\cal B} = -  \sigma \Sigma_{p-1}\int dt dr {\cal L}~,
\nonumber
\eeq
with
\beq
{\cal L} = \left(r \sin\theta \right)^{p-1}\sqrt{1 + r^2  f(r)
\theta_{r}^2 - r^2  f^{-1}(r) \theta_{t}^2}~. 
\nonumber
\eeq
The case with $d=5$ and $p=3$ corresponds to a co-dimension one brane in a five
dimensional bulk, whereas that with $d=4$ and $p=1$ 
corresponds to a string in a four
dimensional spacetime. From the above action the equation of motion
can be obtained and it has to be supplemented by appropriate boundary
conditions. As we have already mentioned we work in the center of mass
frame of the black hole, and the fact that 
the intersection point between the black
hole and the brane is not allowed to move fixes the boundary conditions
at the horizon. The recoil of the black hole in this frame 
is described by giving some initial asymptotically uniform 
velocity to the brane. 
Pictorially the situation is described in {\it
fig}.~\ref{motion}. 
The explicit form of the initial velocity we gave in our simulation 
is $u=v\left(1-{1\over r}\right)$, but the result does not depend much on our choice of
the initial velocity profile. 

\begin{figure}
\scalebox{0.5} {\includegraphics{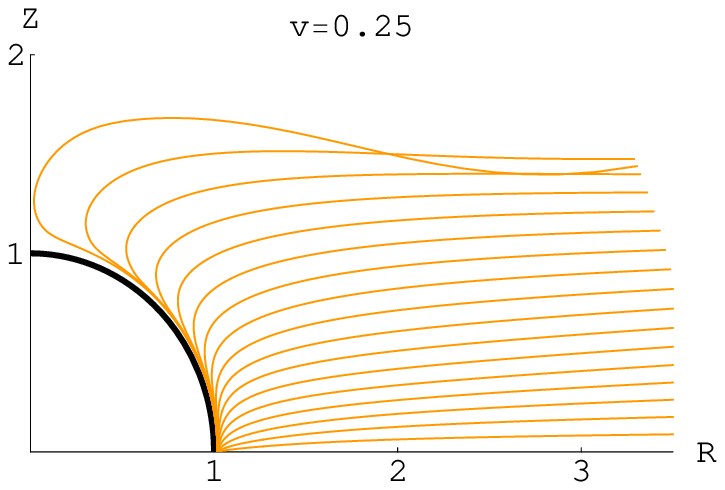}}
\scalebox{0.5} {\includegraphics{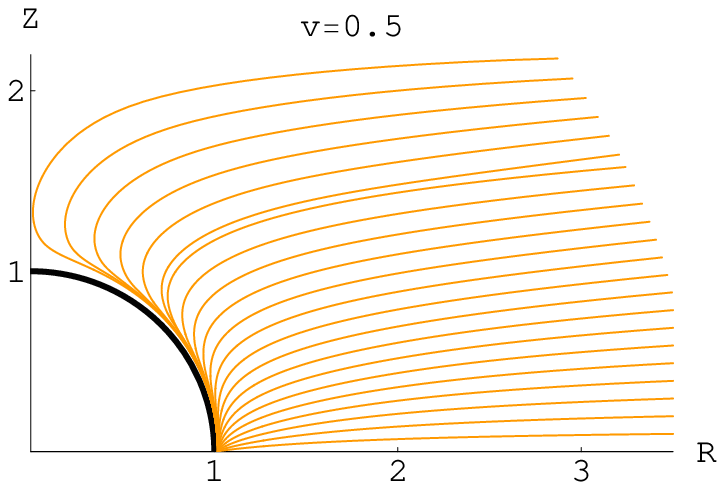}}
\scalebox{0.5} {\includegraphics{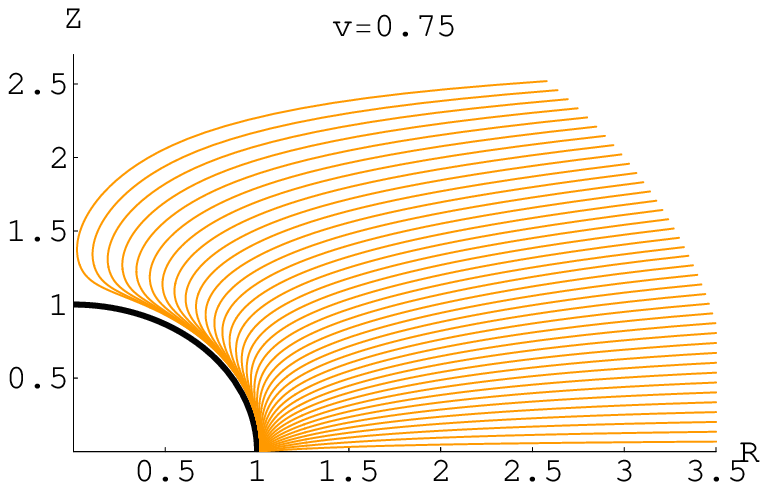}}
\scalebox{0.5} {\includegraphics{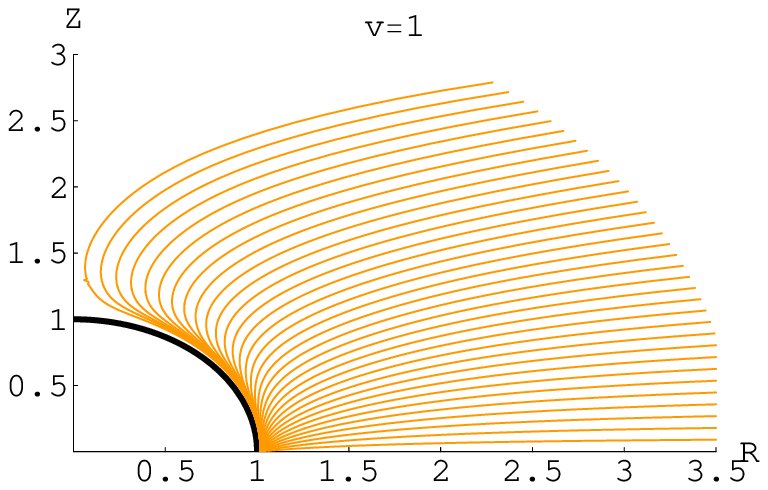}}
\caption{Unstable deformations of the brane: results of the simulation.
 $v$ labels the initial velocity. All the simulations reported refer to a five dimensional bulk and 3+1 dimensional brane and 
the plots are drown in cylindrical coordinates $R=r \sin \theta$ and $Z=-r\cos \theta$.}
\label{simulation}
\end{figure}

The task at hand is to solve the dynamical evolution of the brane. 
This has been done first by writing the equation of motion in a new radial coordinate $\tilde{r}$, defined via the relation $r = 1/(1-e^{-\tilde{r}})$, and then by
applying finite difference methods
along with the Newton-Raphson algorithm to solve the associated non-linear algebraic system. The grid spacing is chosen to be $\delta \tilde{r}= 0.025$ along the
radial direction 
and $\delta t = 0.0025$ along the time direction. 
The number of points of the grid is fixed to be $N_{\tilde{r}} = 134$ along $\tilde{r}$ and varied along $t$ 
depending on the initial velocity. The chosen values are $N_t = 3400, 2420, 1970, 1700$ corresponding to $v=0.25,~0.50,~0.75,~1$ respectively. 
The results of the simulation are reported in {\it fig}.~\ref{simulation}. 
We find that in all cases the brane is bent and eventually the radius 
of the pinched part goes to zero. 
This result does not contradict with the stability analysis 
presented in Ref.~\cite{ishibashi}, in which it was shown 
that there is no linear instability for the case with $d=4$ and $p=2$,  
since the imposed asymptotic boundary conditions are different.  
Roughly speaking, 
the time scale of the escape of a 
black hole is found to be given by $\tau_*\approx R_s/v$ where 
$v$ is the initial recoil velocity. 
In the limit $v\ll 1$ back reaction due to the tension of the brane, 
which is neglected in our present calculation, may 
affect the evolution significantly. 
In this limit, however, the evolution will be well 
approximated by a sequence of static configurations. 
For the case with $d=4$ and $p=2$, static solutions are  
given in Ref.~\cite{cfl}.  
If we imagine the situation that there is an edge of the brane at 
a large radius $r=r_{\rm edge}$, the force necessary  
to sustain the static configuration will be estimated by 
$F=\sigma\Sigma_{p-1}r^{p-1} d(r\theta)/dr|_{r=r_{\rm edge}}$. 
For $p\geq 3$ this force becomes negative when $\theta$ is positive. 
This means that the force between a black hole and the brane is 
repulsive in this static limit. Therefore the effect of the brane 
tension will not prevent the black hole escaping from the brane. 

\begin{figure}
\scalebox{0.5} {\includegraphics{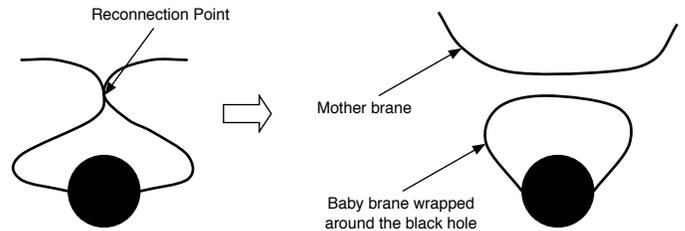}}
\caption{Creation of a baby brane wrapped around the black hole.}
\label{brane_nucleation}
\end{figure}

\noindent
{\it Concluding remarks}:
We have considered a system consisting of a brane plus a black hole. 
We assumed that the size of the black hole is small compared to the extra
dimensions and that the tension of the brane is negligible. In this
approximation, the spacetime with a black hole is well described by
asymptotically flat solution as given in 
Refs.~\cite{tangherlini,mp}. 
If ${\mathcal Z}_2$-symmetry is not imposed, 
it seems natural to consider the situation in
which the black hole acquires some initial velocity with respect to the
brane, say, due to anisotropic emission of particles. 
We simulated this process by studying the dynamical evolution of a 
brane in a fixed black hole spacetime. 
The main result of our study is that, irrespectively of the initial
velocity $v$, 
the brane tends to wrap around the black hole, suggesting that the black hole
might escape in the extra dimensions after two portions of the brane
come in contact and reconnect (see {\it
fig}.~\ref{brane_nucleation}). Such a set-up is of relevance in a number
of physical situations. 

The first application we have in mind is that of accelerator generated
black holes: will it be possible to observe the black hole in its final
state and the
process of evaporation or will the black hole disappear in the extra
dimensions? 
The results of our analysis strongly suggests 
that the black hole will indeed escape from the brane. 
In fact, this process is likely to occur before 
the black hole evaporation completes if the initial mass of 
the produced black hole is sufficiently large. 
The time scale of Hawking evaporation will be estimated as 
$\tau\approx (dM/Mdt)^{-1}\approx M^{-1}(m/M)^{n+3/n+1}$. 
On the other hand, 
the time scale of the escape of a 
black hole was found to be given by $\tau_*\approx R_s/v$. 
If the origin of the recoil 
is the inhomogeneous emission of particles due to 
Hawking radiation, we will have a rough estimate 
$v\approx \alpha/\sqrt{N}$, where $\alpha$ is 
the fraction of the evaporated mass and $N\approx \alpha M/T_H$ 
is the number of emitted particles. 
Then we have $\tau_*\approx M^{-1}(m/M)^{n+4\over 2(n+1)}
/\sqrt{\alpha}$. 
For $\alpha=O(1)$ and $m\agt M$, 
the time scale of escape is shorter than the time scale of 
evaporation. 
Furthermore, $\alpha$ at $\tau_*$ is estimated as 
$\alpha \approx \tau_*/\tau$. Using this relation, 
we find $\tau_*\approx M^{-1}(m/M)^{2n+7\over 3(n+1)}$. 

Aside from more phenomenological application mentioned above, it is
amusing to speculate about other possible situations where the process
of baby brane nucleation may be relevant. An interesting application is
the possibility to generate global charge non-conserving processes on the
defect. This has already been noticed in Ref.~\cite{dg}, where the
quantum fluctuations of the geometry induce nucleation of baby branes
and this is used as a mechanism to generate baryon asymmetry in the
early universe. Our example, though, is different in that it is entirely
classical and provides an explicit example of how this nucleation takes
place.
All this is also reminiscent of the baryogenesis mechanism proposed in
Refs.~\cite{hawking,zeldovich}, where the evaporation of primordial black
hole is used as a means
to generate the asymmetry. Of course a more detailed study is necessary
to quantify whether or not such a possibility is feasible
\cite{nino-takahiro-forth}.

One can push things even further by relaxing the condition that the
fundamental Planck scale is in the TeV range, but still assume a higher
dimensional spacetime. This will not change the process of baby brane
creation but will reduce the size of such objects from $10^{-16}$ cm to the traditional
Planck  length $l_P \sim 10^{-33}$ cm. It is, then, tempting to imagine the spacetime at
the Planck scale
as a foam of baby branes and the extra dimensions filled with such
bubbles \cite{nino-takahiro-forth}.

In the present paper, we completely neglected the self-gravity of the 
brane. When ${\mathcal Z}_2$-symmetry across the brane is imposed, 
the brane bending is not allowed in our current treatment. 
Once the self-gravity of the brane is turned on, however, it is not 
completely clear whether the escape of black holes from the brane is 
forbidden or not. In fact, such a process is 
thought to be one possible mechanism to explain the 
proposed conjecture of classical black hole evaporation  
in the Randall-Sundrum II model~\cite{takahiro, kaloper}. 
This is a challenging problem which requires development of 
numerical relativity with a self-gravitating 
singular hypersurface and a black hole. 

%%%%%%%%%%%%%%%%%%%%%%%%%%%%%%%%
%\section*{Acknowledgements}
We wish to thank C. Germani, O. Pujol\`as and M. Sasaki for useful discussions.
This work is supported in part by 
Grant-in-Aid for Scientific Research, No. 16740165 from Japan Society for Promotion of Science 
and by that for the 21st Century COE
``Center for Diversity and Universality in Physics'' at 
Kyoto university, both from the Ministry of
Education, Culture, Sports, Science and Technology of Japan. 
A.F. is supported by the JSPS under contract No. P047724.

%%%%%%%%%%%%%%%%%%%%%%%%%%%%%%%%

%%%%%%%%%%%%%%%%%%%%%%%%%%%

\end{document}